\documentclass[aps, prl, twocolumn, showpacs, superscriptaddress]{revtex4-1}
\usepackage{main}
\usepackage{amssymb}

\begin{document}

    \title{Manipulating Andreev and Majorana Bound States with microwaves}

    \spsmsAuthor{Joseph Weston}
    \spsmsAuthor{Benoit Gaury}
    \spsmsAuthor{Xavier Waintal}
    \date{\today}

\begin{abstract}
We study the interplay between Andreev (Majorana) bound states that form at the boundary of a
(topological) superconductor and a train of microwave pulses. We find that the
extra dynamical phase coming from the pulses can shift the phase of the Andreev reflection,
resulting in the appearance of dynamical Andreev states.
As an application we study the presence of the zero bias peak in the differential conductance of
a normal-topological superconductor junction - the simplest, yet somehow ambiguous, experimental signature for Majorana
states.
Adding microwave radiation to the measuring electrodes acts as a probe of the Andreev
nature of the zero bias peak.
\end{abstract}

\maketitle

Andreev bound states and their topological counterparts, Majorana states,
have seen  considerable renewed  interest springing from the
possibility of using the latter for topologically protected quantum
computation~\cite{nayak_non-abelian_2008}. These Majorana states were first proposed to appear
at the boundaries of so-called topological superconductors (TSs), rather exotic
materials~\cite{kitaev_unpaired_2001,ivanov_non-abelian_2001,fu_superconducting_2008}.  Although the search for a ``natural'' TS is very active, another route consists of engineering a TS by putting a
regular (s-wave) superconductor into contact with another material that has strong
spin-orbit interaction. A quantum spin Hall (QSH) topological insulator in contact
with a superconductor is a popular possibility~\cite{mi_proposal_2013}.  Another
promising proposal is to use a semiconducting nanowire in contact with a
superconductor~\cite{lutchyn_majorana_2010, oreg_helical_2010}. Indeed, a
possible signature of the existence of Majorana states in a semi-conducting nanowire
coupled to a superconducting electrode was recently reported~\cite{das_zero-bias_2012, mourik_signatures_2012,
deng_anomalous_2012,churchill_superconductor-nanowire_2013}.

In this letter we report on a dynamical generalization of Andreev/Majorana states in the
presence of train of microwave pulses, i.e. the possibility to detect and manipulate these
states by bringing the system out of equilibrium. Our proposal shares some features
with other recent proposals that went under the name of ``Floquet Majorana''
states~\cite{jiang_majorana_2011,liu_topological_2012,liu_floquet_2013, kundu_transport_2013}: zero energy
bound states that can be stabilized by a finite frequency perturbation. Our mechanism is, however, rather different
and uses a simpler perturbation (oscillating voltage) applied to an electrode (as opposed to the whole system)
together with much smaller frequencies (\si{\GHz} to \si{\THz}).

The dynamical Andreev states could be used to discriminate true Majoranas from other phenomena.
 Indeed, experimental evidence reported so far for Majorana states are based entirely on the presence (or lack) of a zero bias peak in the differential conductance. The peak could, however, originate from different
sources~\cite{pikulin_zero-voltage_2012,lee_zero-bias_2012,chang_tunneling_2013,
cheng_interplay_2014,zitko_shiba_2015,liu_zero-bias_2012,lutchyn_search_2011,kells_low-energy_2012,
lee_spin-resolved_2014,rieder_endstates_2012,
stanescu_nonlocality_2014,kells_near-zero-energy_2012,roy_topologically_2013};
the interpretation of the experiments so far is still debated. Here we propose
to measure the (DC) differential conductance in the presence of a periodic train of
voltage pulses. Our findings indicate that such a probe provides a way of performing
spectroscopy on the bound states, permitting an identification of
conductance peaks as Andreev states.

\begin{figure}
   \includegraphics[width=0.44\textwidth]{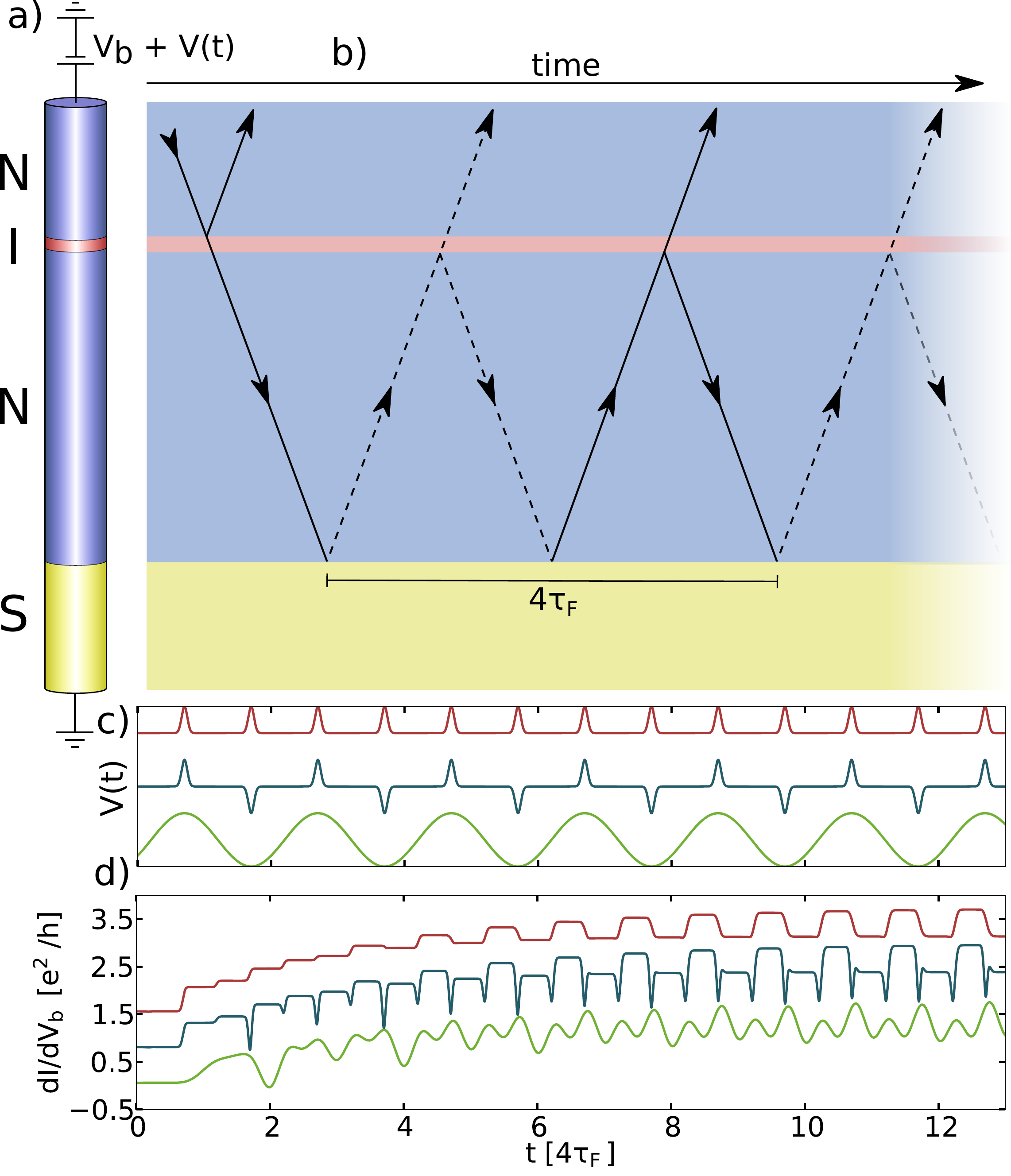}
   \caption{a) Sketch of the setup: blue is normal conductor, red is
            insulating barrier and yellow is superconductor.
            b) Sketch of the different trajectories for electrons (full lines) and holes (dashed).  
            c) Different applied voltage pulse trains as a function of time.
            From top to bottom: upright pulses, alternated pulses and sine pulses.
            d) Numerical computation of the zero bias differential
            conductance as a function of time for different pulse trains.
            (with vertical offset for clarity.)
            }
            \label{fig:setup}
\end{figure}

\paragraph{Scattering approach to Andreev and Majorana states.}
To be specific, we focus
on the setup shown in \figref{fig:setup}a): a nanowire
connected to a superconducting electrode (S, yellow) on one side and to a normal
electrode on the other side (N, blue). A short insulating barrier (I, red) (created
from either a gate or the Schottky barrier at the
semiconductor-electrode interface) serves to confine possible bound states. 
Without the barrier, these states would hybridize
strongly with the continuum. A weak hybridization
transforms the bound states into resonances and allows one to probe them
by measuring the DC differential conductance $dI/dV_b$ through the wire.
The superconducting electrode is grounded, while a bias voltage $V_b+V(t)$ is
applied to the normal electrode. $V(t)$ is periodic with period $T$.

\begin{figure}
\includegraphics[width=0.44\textwidth]{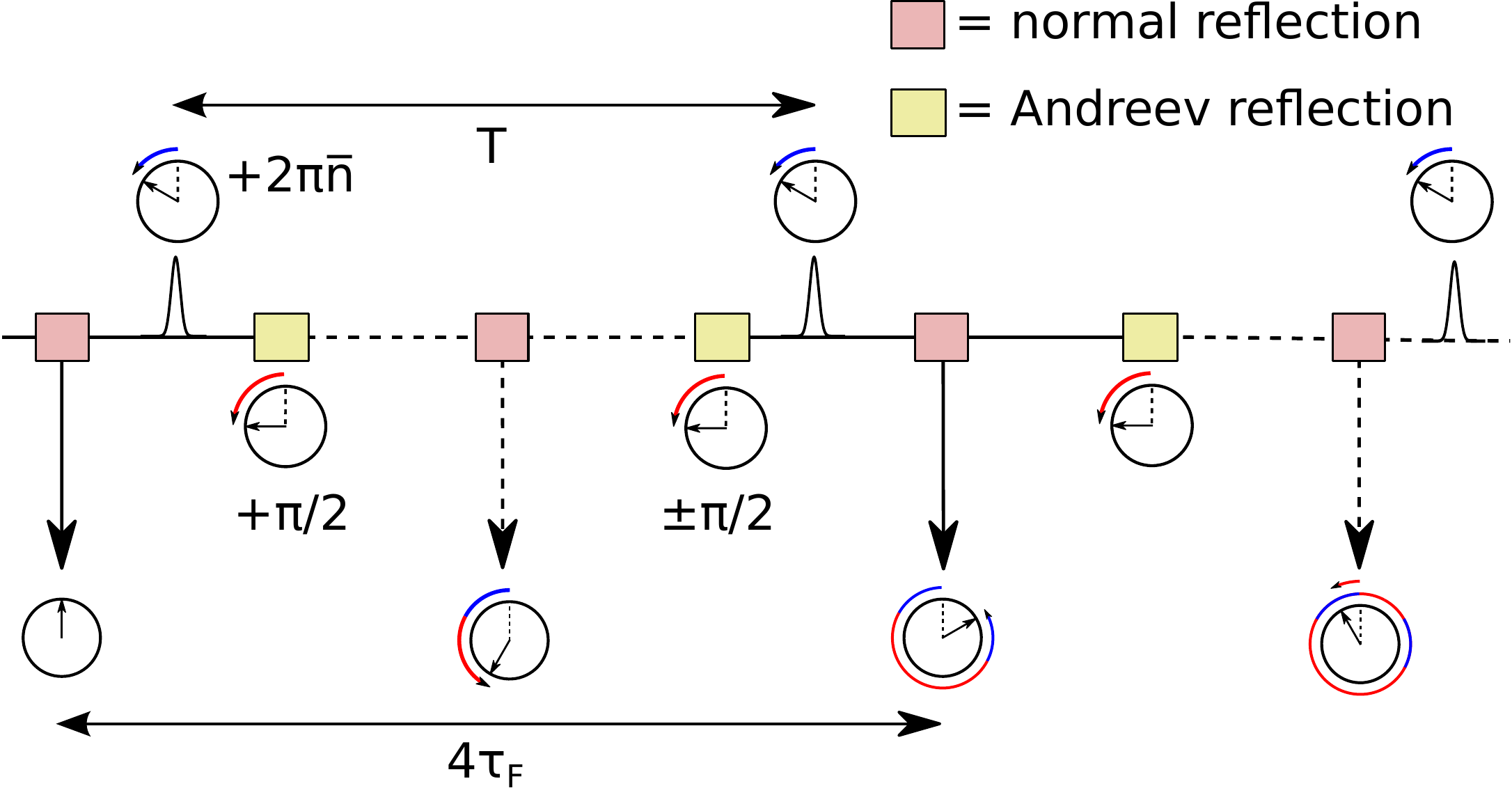}
\caption{Sketch of the accumulated phase along the paths through the
interferometer in the resonant situation $T=4\tau_F$. The paths have been flattened out for clarity. The clocks show
the total phase accumulated (bottom) and the phase contributions from the
pulses (blue arrows) and Andreev reflection (red arrows).
}
\label{fig:paths}
\end{figure}
Scattering theory provides a very intuitive way for understanding Andreev
states in terms of an effective Fabry-Perot cavity. Following
the approach developed in Ref.~\onlinecite{mi_proposal_2013}, an electron
injected from the normal electrode undergoes transmission (reflection) with
amplitude $d$ ($r$) at the barrier and Andreev reflection (an electron is
reflected as a hole) at the normal-superconducting interface.
The total amplitude for an electron injected from the normal contact being reflected
into the contact as a hole (with creation of a Cooper pair inside the superconductor) is given
by summing over the different trajectories sketched in \figref{fig:setup}b),
\begin{equation}
    r_{he}(E) = d^2 r_A \sum_{m=0}^{\infty} r^{2m}r_A^{2m} z^{4m}.
\label{eq:amplitude}
\end{equation}
The energy $E$ is measured from the Fermi energy, $E_F$. When $E$ is smaller than
the superconducting gap $\Delta$, $r_A$ is a pure phase
given by $r_A^2=(-1)^{Z_2} e^{-2i\text{arcos}(E/\Delta)}$
where $Z_2=0$ for a conventional superconductor and $Z_2=1$ for a topological
one~\cite{beenakker_random-matrix_2014}. The factor $z$ accounts for the phase accumulated in
the normal region, $z=e^{iE \tau_F/\hbar}$, where $\tau_F$ is the time of flight
between the superconducting interface and the barrier. The geometric series
\eqref{eq:amplitude} can be readily resummed into $r_{he}(E) = d^2 r_A/
(1-r^2 e^{\Phi(E)})$ with,
\begin{equation}
\Phi(E)= 2\text{arcos}(E/\Delta) + \frac{4E\tau_F}{\hbar} +\pi Z_2
\end{equation}
When the insulating barrier is very high (perfect reflection $r=1$) one observes
true bound states in the spacer between I and S with energies $E_p$ satisfying
$\Phi(E_p)=2\pi p$ with $p=0,1,2\dots$. The zero energy state $E_1 = 0$ present
for TS is the Majorana bound state~\cite{law_majorana_2009, mi_proposal_2013}. Upon increasing the transparency of the insulating barrier,
the Andreev states hybridize with the conducting electrode allowing one to probe them through
the differential conductance $dI/dV_b=(4e^2/h)/(1+Z_2)|r_{eh}(eV_b)|^2$ of the system (note that the TS is
spinless, hence the factor 1/2).

The theory for the time-dependent current in the presence of an additional train of
pulses follows a very similar line. We first focus on the
long junction limit where there are many resonances inside the superconducting gap,
$h/\tau_F \ll \Delta$. This condition will be relaxed in the numerical computations.
Following Refs.~\onlinecite{gaury_numerical_2014,gaury_dynamical_2014} we find the
time-dependent electron-hole reflection amplitude to be,
\begin{equation}
            r_{he}(t,E) = d^2 r_A \sum_{m=0}^{\infty} r^{2m}e^{i\Phi_m(t,E)}
            \label{eq:amplitude_t}
\end{equation}
with
 \begin{equation}
\Phi_m(t,E)= m\Phi(E) +\phi(t + 4\tau_Fm)
\end{equation}
where $\phi(t) = (e/h) \int_0^tV(u)\,du$ is the additional phase due to the
time-dependent potential $V(t)$. This phase is translated by
$4\tau_Fm$ for different trajectories to account for the different times of
propagation from the normal contact. It is this phase that will be used to
manipulate the resonances.
We will consider trains of pulses of different shapes
as shown in \figref{fig:setup}c).
We denote $2\pi\bar{n}$ the phase  accumulated over one period $\phi(T)$ (upright pulses)
or half a period $\phi(T/2)$(alternating and sine pulses).
In analogy with the DC case, the time-dependent differential conductance is,
\begin{equation}
    \frac{dI(t)}{dV_b}=\frac{4e^2}{h(1+Z_2)} |r_{eh}(t,eV_b)|^2
    \label{eq:dIdV}
\end{equation}

\begin{figure*}[t]
    \includegraphics[width=\textwidth]{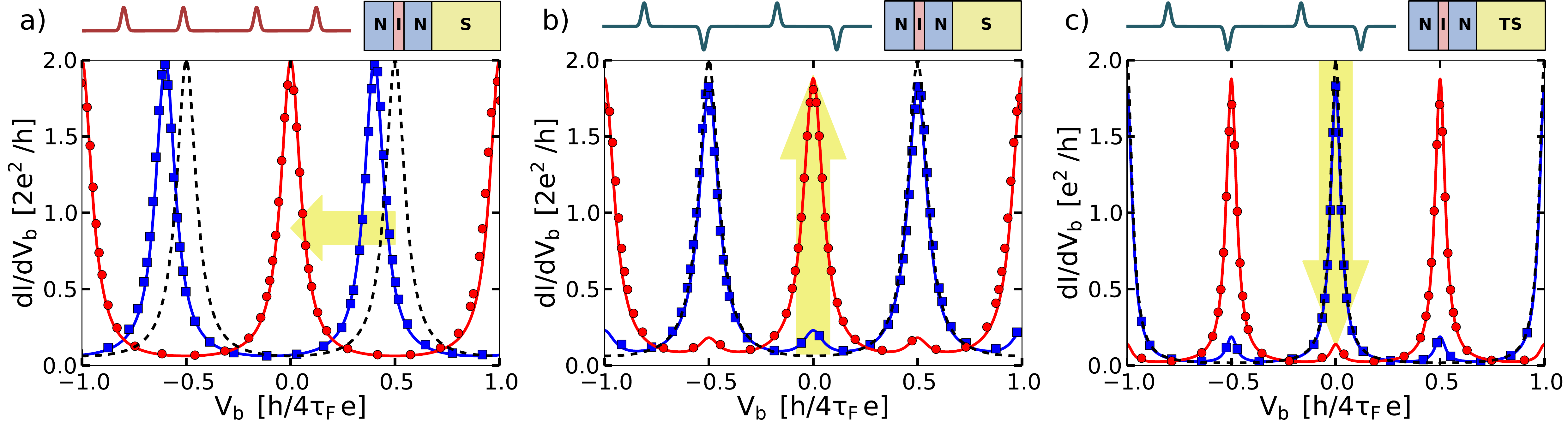}
\caption{DC conductance in the presence of a train of voltage pulses
    with small intensity $\bar{n} = 0.1$ (blue squares) and large intensity $\bar{n} = 0.5$
    (red circles). Dashed line: no voltage pulse, 
    Symbols: numerical simulations, lines: analytical model, Yellow arrows: evolution with increasing $\bar{n}$.
    The figure above each plot shows which type of voltage pulse train and
    system (regular $S$ or topological superconductor $TS$) was used. For the regular
    system we used: $\alpha=0$, $\Delta=0.075$, $E_F=2.0$,
    $E_z=0$ and $V_t=3.0$ (giving a barrier transmission of $|d|^2=0.3$).
    For the topological system the parameters used were: $\alpha=1.0$,
    $\Delta=0.5$, $E_F=1.0$, $E_z=2.0$ and $V_t=5.0$ (giving a barrier
    transmission of $|d|^2=0.17$).
}
\label{fig:G_Vb}
\end{figure*}

The simplest situation is when one sends a series of upright localized pulses (of widths
much shorter than $\tau_F$).  \figref{fig:paths} sketches the phase accumulated
along the different paths through the junction; it corresponds to a flattened
description of the paths shown in \figref{fig:setup}b).  Whenever one crosses
a pulse one picks up a phase $2\pi \bar n$, while a phase $\pi/2$ is picked up
upon Andreev reflection at the NS interface (for conventional
superconductors;  for topological ones, one alternately picks up $\pi/2$ and
$-\pi/2$, \cite{pikulin_zero-voltage_2012}).  When the period of the pulse train
exactly matches the delay between different trajectories, $T=4\tau_F$, the
phase $\phi(t+4m\tau_F)$ is simply given by $\phi(t+4m\tau_F)= m (2\pi \bar
n)$.  As a result, $\Phi_m(t,E)=m \Phi(E) + m (2\pi\bar n)$ and
\eqref{eq:dIdV} takes the form of a geometric series, i.e.
{\it the (only) effect of the pulse is to shift the resonance by $2\pi\bar n$}:
\begin{equation}
r_{he}(E) = \frac{d^2 r_A}{1-r^2 e^{i\Phi(E)+2i\pi\bar n }}
\end{equation}
  An example of differential conductance for increasing $\bar n$  is
given in \figref{fig:G_Vb}a).  In particular, for $\bar n=1/2$ one transforms
the spectrum of a conventional superconductor into the spectrum of a TS
and vice versa.  Also, for $\bar n=1/2$, the train of pulses is equivalent to a
train of alternating pulses; \figref{fig:G_Vb}b) corresponds to this alternating train. We see that upon
increasing $\bar n$ (using the amplitude of the pulses), a zero-bias peak develops while the
Andreev peaks that were initially present shrink. \figref{fig:G_Vb}c) shows the
effect of a similar alternating train on a system  in the
topological phase. We see that the initial zero-bias (Majorana) peak is
effectively destroyed when $\bar{n} = 0.5$ while new Andreev peaks appear; the
dual to the previous situation. Anticipating what follows, we find that a
simple monochromatic sine pulse $\propto \cos \omega t$ (which can be seen as a
distortion of the alternating pulse train) will have a similar, although
slightly less marked, effect.

To proceed with the generic case where $T\ne 4\tau_F$, we expand $\phi(t)$ as a
Fourier series, $e^{i\phi(t)}$\:$=$\:$e^{i\phi(T)t/T}\sum_{p=-\infty}^{+\infty}
c_p e^{ip \omega t}$.
For a simple monochromatic pulse $V(t)=V_0 \cos \omega t$ , the Fourier coefficients are given by
$c_p=J_p(eV_0/\hbar\omega)$ where $J_p(x)$ is the $p^{th}$ Bessel function of the first kind.
Keeping only the DC part of the current, we get
\begin{equation}
\frac{dI}{dV_b}=\frac{4e^2|d|^4}{h(1+Z_2)}
            \sum_{p=-\infty}^{+\infty}
            \left|
            \frac{c_p}{1 - |r|^2
            e^{i\Phi(eV_b) + 4i\omega\tau_F (\phi(T)/2\pi + p)}}
            \right|^2
\label{eq:probability}
\end{equation}
The appearance of the resonances in \eqref{eq:probability} corresponds to
minimizing the denominator, i.e. the phase in the denominator is a
multiple of $2\pi$.  We thus expect (for $eV_b\ll\Delta$) sharp peaks
at positions given by
\begin{equation}
\frac{eV_b}{h} + \frac{\omega}{2\pi}(p\,+\,\phi(T)/2\pi) =
\frac{1}{4\tau_F} \left[q
- \frac{1}{2} (1+Z_2) \right] \quad p,q \in \mathbb{Z}
\label{eq:condition}
\end{equation}

\begin{figure*}[t]
    \includegraphics[width=0.9\textwidth]{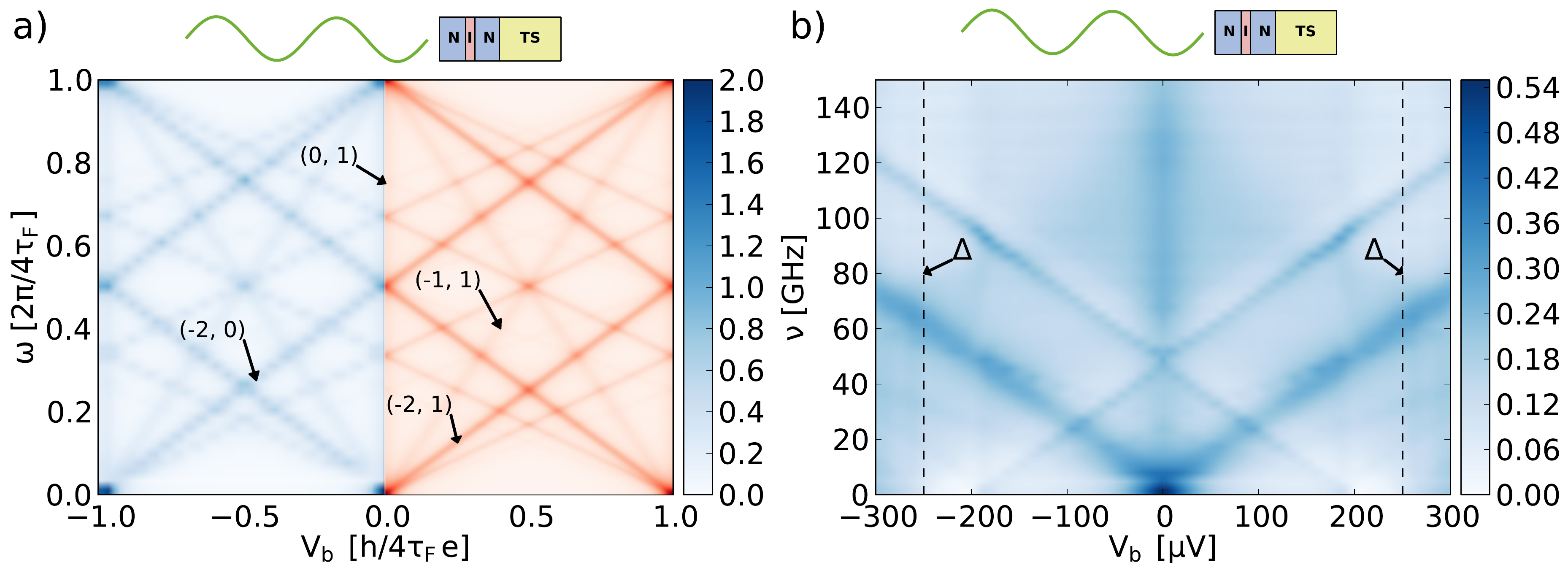}
    \caption{Topological junction. Conductance in the presence of a sinusoidal perturbation as a function of DC bias,
        $V_b$, and driving frequency, $\omega$ for $\bar n = 0.5$. The color scales are given in
        units of $e^2/h$. 
        a)~Long junction. Negative (positive) $V_b$ shows the results from microscopic
        numerical (analytical with \eqref{eq:probability}) calculations. Some of the principle contributions for different pairs
        $(p, q)$ (\eqref{eq:condition}) are marked. $|d|^2 = 0.17$.
        b)~Shorter junction, obtained from numerical calculations including static disorder (quasi-ballistic regime)
        and at a finite temperature of \SI{20}{\milli\kelvin} (for details see the supplementary material).
    }
    \label{fig:colorplots}
\end{figure*}

 \figref{fig:colorplots} shows colorplots of the differential conductance
for the system in the topological regime in the presence of a sinusoidal time-dependent
voltage. The different lines visible on \figref{fig:colorplots}a) correspond to different
pairs of the integers $p$ and $q$ of \eqref{eq:condition}. It now becomes clear that the presence of the
resonant peaks with conductance $2e^2/h$ occur when lines for all values of $p$
coincide. Resonant peaks with smaller conductance values occur where subsets of
the lines coincide. One can understand each of the terms
in \eqref{eq:probability} as referring to a $p$-photon absorption process for the $q^{th}$ resonance. The presence of this characteristic structure in the transport measurements could
be used to identify 0-bias peaks as being Andreev/Majorana bound states -- as opposed to Kondo
resonances or Shiba states, as proposed in Ref.~\onlinecite{lee_zero-bias_2012} --
as the frequency dependence is closely linked to the time of flight of the resonant
cavity, i.e. to the Fabry-Perot nature of the Majorana resonance.

\paragraph{Microscopic (numerical) calculations.}
To go beyond the above analytical calculation, we now turn to a 
microscopic approach following the proposal
made in Refs.~\cite{chevallier_andreev_2013,oreg_helical_2010}. We
consider a one dimensional semiconductor wire with (strong) spin-orbit
interaction and an external magnetic field. The combination of these two
ingredients makes the system effectively spinless close to the Fermi level, so
that a nearby (s-wave) superconductor induces p-wave topological
superconductivity, and the presence of Majorana fermions in a certain range of
parameters (see the supplementary material). The Hamiltonian reads,
$H = \int dy \Psi^\dagger(y)\mathcal{H}(y)\Psi(y)$ with
$\Psi^\dagger(y) = (\psi^\dagger_\uparrow(y),
 \psi^\dagger_\downarrow(y), \psi_\downarrow(y), -\psi_\uparrow(y))$
where $\psi^{(\dagger)}_{\uparrow(\downarrow)}(y)$ are the electron
annihilation (creation) operators with spin up (down) at position $y$.
The Effective Bogoliubov-de~Gennes  Hamiltonian reads,
\begin{align}
\begin{split}
\mathcal{H}= &\left(\frac{p^2}{2m^*} + \alpha p\sigma_x +
V_t\delta(y) + V(t)\Theta(-y-L_N) - E_F\right)\tau_z
\\&+ U(y)\tau_z + \Delta\Theta(y - L_S)\tau_x + E_z\sigma_z
\label{eq:hamiltonian}
\end{split}
\end{align}
where $p=i\hbar\partial/\partial y$ is the momentum operator, $m^*$ is the effective mass,
$E_z$ is the Zeeman splitting, $\alpha$ is the Rashba spin-orbit interaction, $V_t$ is the strength
of the tunneling barrier and $E_F$ is the Fermi energy.  $\Theta(y)$
is the Heaviside step function. The $\sigma_a$ and $\tau_a$ ($a=x,y,z$) are Pauli matrices operating on the spin and particle-hole subspaces respectively. $U(y)$ is an
on site (Anderson) disorder potential present in the normal part of the wire. The wire is
subject to a time-dependent voltage perturbation $V(t)$ applied to the normal
reservoir (at $y \le -L_N$) where we assume that the voltage drop is
sharp~\cite{gaury_dynamical_2014}. The superconductor of gap $\Delta$ is placed
at $y \ge L_S$. 
The time-dependent perturbation is switched on at $t=0$.
After a transient regime (of approximate duration $40\tau_F$) 
the system relaxes to a periodic regime, see \figref{fig:setup}. The DC differential conductance corresponds to the
average over one period of the traces shown in \figref{fig:setup}d) in the periodic regime. Simulations
were carried out using systems discretized on \numrange{400}{3200} sites
and for times up to $144000\ \hbar/E_F$ with the help of the technique described in Ref.~\cite{gaury_numerical_2014}
and of the Kwant package~\cite{groth_kwant:_2014}

We first focus on the simple long junction regime of the analytics. Perfect Andreev reflection, implies 
$\Delta \ll E_F$ 
(we use $\Delta = 0.17 E_F$ in the simulations) while the long junction limit corresponds to $h/\tau_F \ll \Delta$ 
(we use $h/\tau_F \sim 0.01 \Delta$ ).
\figref{fig:G_Vb} [as well as \figref{fig:colorplots}a)] shows a perfect agreement between the simulations (symbols) and the
analytical result (lines). 

Let us now turn to a regime closer to the experiments performed in the
group of Kouwenhoven\cite{mourik_signatures_2012}. This implies three important modifications to the above picture. First, the Thouless energy
$\hbar/\tau_F$ is of the same order as the induced gap $\Delta$ in the wire. This is relevant as upon applying the radio-frequency field, many quasi-particles will be excited in the superconductor and one anticipates a corresponding decrease in the amplitude of Andreev reflection. Second, we add disorder to the simulation, corresponding to a mean free path of the same size as the length of the wire (i.e. the system is in the crossover between the ballistic and diffusive regime). Third, the simulations are performed at finite (typical dilution fridge) temperature. These results are shown in
\figref{fig:colorplots}b) (see also the supplementary material). We find that the key features of the theory are robust with respect to these perturbations: although the resonances are a bit wider and the thinnest ones have disappeared, the main ones remain perfectly visible. The frequencies required to perform this experiment are within reach of current commercially available apparatus.

\paragraph{Conclusion.} We have shown that the interplay between a train of pulses and
Andreev/Majorana states can provide a differential conductance identical to the DC case
except for a tunable dynamical phase (upright pulses) as well as a spectroscopy of the
states (sine pulses). Beyond the present study, we anticipate that the same line of thought
could be used for more elaborate manipulations such as those needed for quantum computation.

\paragraph{Acknowledgements}This work was supported by the ERC grant MESOQMC from the
European Union. We thank M. Houzet and J. Meyer for interesting discussions.

\bibliographystyle{apsrev}
\bibliography{references}

\begin{thebibliography}{34}
\expandafter\ifx\csname natexlab\endcsname\relax\def\natexlab#1{#1}\fi
\expandafter\ifx\csname bibnamefont\endcsname\relax
  \def\bibnamefont#1{#1}\fi
\expandafter\ifx\csname bibfnamefont\endcsname\relax
  \def\bibfnamefont#1{#1}\fi
\expandafter\ifx\csname citenamefont\endcsname\relax
  \def\citenamefont#1{#1}\fi
\expandafter\ifx\csname url\endcsname\relax
  \def\url#1{\texttt{#1}}\fi
\expandafter\ifx\csname urlprefix\endcsname\relax\def\urlprefix{URL }\fi
\providecommand{\bibinfo}[2]{#2}
\providecommand{\eprint}[2][]{\url{#2}}

\bibitem[{\citenamefont{Nayak et~al.}(2008)\citenamefont{Nayak, Simon, Stern,
  Freedman, and Das~Sarma}}]{nayak_non-abelian_2008}
\bibinfo{author}{\bibfnamefont{C.}~\bibnamefont{Nayak}},
  \bibinfo{author}{\bibfnamefont{S.~H.} \bibnamefont{Simon}},
  \bibinfo{author}{\bibfnamefont{A.}~\bibnamefont{Stern}},
  \bibinfo{author}{\bibfnamefont{M.}~\bibnamefont{Freedman}}, \bibnamefont{and}
  \bibinfo{author}{\bibfnamefont{S.}~\bibnamefont{Das~Sarma}},
  \bibinfo{journal}{Rev. Mod. Phys.} \textbf{\bibinfo{volume}{80}},
  \bibinfo{pages}{1083} (\bibinfo{year}{2008}).

\bibitem[{\citenamefont{Kitaev}(2001)}]{kitaev_unpaired_2001}
\bibinfo{author}{\bibfnamefont{A.~Y.} \bibnamefont{Kitaev}},
  \bibinfo{journal}{Phys.-Usp.} \textbf{\bibinfo{volume}{44}},
  \bibinfo{pages}{131} (\bibinfo{year}{2001}), .

\bibitem[{\citenamefont{Ivanov}(2001)}]{ivanov_non-abelian_2001}
\bibinfo{author}{\bibfnamefont{D.~A.} \bibnamefont{Ivanov}},
  \bibinfo{journal}{Phys. Rev. Lett.} \textbf{\bibinfo{volume}{86}},
  \bibinfo{pages}{268} (\bibinfo{year}{2001}).

\bibitem[{\citenamefont{Fu and Kane}(2008)}]{fu_superconducting_2008}
\bibinfo{author}{\bibfnamefont{L.}~\bibnamefont{Fu}} \bibnamefont{and}
  \bibinfo{author}{\bibfnamefont{C.~L.} \bibnamefont{Kane}},
  \bibinfo{journal}{Phys. Rev. Lett.} \textbf{\bibinfo{volume}{100}},
  \bibinfo{pages}{096407} (\bibinfo{year}{2008}).

\bibitem[{\citenamefont{Mi et~al.}(2013)\citenamefont{Mi, Pikulin, Wimmer, and
  Beenakker}}]{mi_proposal_2013}
\bibinfo{author}{\bibfnamefont{S.}~\bibnamefont{Mi}},
  \bibinfo{author}{\bibfnamefont{D.~I.} \bibnamefont{Pikulin}},
  \bibinfo{author}{\bibfnamefont{M.}~\bibnamefont{Wimmer}}, \bibnamefont{and}
  \bibinfo{author}{\bibfnamefont{C.~W.~J.} \bibnamefont{Beenakker}},
  \bibinfo{journal}{Phys. Rev. B} \textbf{\bibinfo{volume}{87}},
  \bibinfo{pages}{241405} (\bibinfo{year}{2013}).

\bibitem[{\citenamefont{Lutchyn et~al.}(2010)\citenamefont{Lutchyn, Sau, and
  Das~Sarma}}]{lutchyn_majorana_2010}
\bibinfo{author}{\bibfnamefont{R.~M.} \bibnamefont{Lutchyn}},
  \bibinfo{author}{\bibfnamefont{J.~D.} \bibnamefont{Sau}}, \bibnamefont{and}
  \bibinfo{author}{\bibfnamefont{S.}~\bibnamefont{Das~Sarma}},
  \bibinfo{journal}{Phys. Rev. Lett.} \textbf{\bibinfo{volume}{105}},
  \bibinfo{pages}{077001} (\bibinfo{year}{2010}).

\bibitem[{\citenamefont{Oreg et~al.}(2010)\citenamefont{Oreg, Refael, and von
  Oppen}}]{oreg_helical_2010}
\bibinfo{author}{\bibfnamefont{Y.}~\bibnamefont{Oreg}},
  \bibinfo{author}{\bibfnamefont{G.}~\bibnamefont{Refael}}, \bibnamefont{and}
  \bibinfo{author}{\bibfnamefont{F.}~\bibnamefont{von Oppen}},
  \bibinfo{journal}{Phys. Rev. Lett.} \textbf{\bibinfo{volume}{105}},
  \bibinfo{pages}{177002} (\bibinfo{year}{2010}).

\bibitem[{\citenamefont{Das et~al.}(2012)\citenamefont{Das, Ronen, Most, Oreg,
  Heiblum, and Shtrikman}}]{das_zero-bias_2012}
\bibinfo{author}{\bibfnamefont{A.}~\bibnamefont{Das}},
  \bibinfo{author}{\bibfnamefont{Y.}~\bibnamefont{Ronen}},
  \bibinfo{author}{\bibfnamefont{Y.}~\bibnamefont{Most}},
  \bibinfo{author}{\bibfnamefont{Y.}~\bibnamefont{Oreg}},
  \bibinfo{author}{\bibfnamefont{M.}~\bibnamefont{Heiblum}}, \bibnamefont{and}
  \bibinfo{author}{\bibfnamefont{H.}~\bibnamefont{Shtrikman}},
  \bibinfo{journal}{Nat Phys} \textbf{\bibinfo{volume}{8}},
  \bibinfo{pages}{887} (\bibinfo{year}{2012}), .

\bibitem[{\citenamefont{Mourik et~al.}(2012)\citenamefont{Mourik, Zuo, Frolov,
  Plissard, Bakkers, and Kouwenhoven}}]{mourik_signatures_2012}
\bibinfo{author}{\bibfnamefont{V.}~\bibnamefont{Mourik}},
  \bibinfo{author}{\bibfnamefont{K.}~\bibnamefont{Zuo}},
  \bibinfo{author}{\bibfnamefont{S.~M.} \bibnamefont{Frolov}},
  \bibinfo{author}{\bibfnamefont{S.~R.} \bibnamefont{Plissard}},
  \bibinfo{author}{\bibfnamefont{E.~P. a.~M.} \bibnamefont{Bakkers}},
  \bibnamefont{and} \bibinfo{author}{\bibfnamefont{L.~P.}
  \bibnamefont{Kouwenhoven}}, \bibinfo{journal}{Science}
  \textbf{\bibinfo{volume}{336}}, \bibinfo{pages}{1003} (\bibinfo{year}{2012}),
  .

\bibitem[{\citenamefont{Deng et~al.}(2012)\citenamefont{Deng, Yu, Huang,
  Larsson, Caroff, and Xu}}]{deng_anomalous_2012}
\bibinfo{author}{\bibfnamefont{M.~T.} \bibnamefont{Deng}},
  \bibinfo{author}{\bibfnamefont{C.~L.} \bibnamefont{Yu}},
  \bibinfo{author}{\bibfnamefont{G.~Y.} \bibnamefont{Huang}},
  \bibinfo{author}{\bibfnamefont{M.}~\bibnamefont{Larsson}},
  \bibinfo{author}{\bibfnamefont{P.}~\bibnamefont{Caroff}}, \bibnamefont{and}
  \bibinfo{author}{\bibfnamefont{H.~Q.} \bibnamefont{Xu}},
  \bibinfo{journal}{Nano Lett.} \textbf{\bibinfo{volume}{12}},
  \bibinfo{pages}{6414} (\bibinfo{year}{2012}), .

\bibitem[{\citenamefont{Churchill et~al.}(2013)\citenamefont{Churchill, Fatemi,
  Grove-Rasmussen, Deng, Caroff, Xu, and
  Marcus}}]{churchill_superconductor-nanowire_2013}
\bibinfo{author}{\bibfnamefont{H.~O.~H.} \bibnamefont{Churchill}},
  \bibinfo{author}{\bibfnamefont{V.}~\bibnamefont{Fatemi}},
  \bibinfo{author}{\bibfnamefont{K.}~\bibnamefont{Grove-Rasmussen}},
  \bibinfo{author}{\bibfnamefont{M.~T.} \bibnamefont{Deng}},
  \bibinfo{author}{\bibfnamefont{P.}~\bibnamefont{Caroff}},
  \bibinfo{author}{\bibfnamefont{H.~Q.} \bibnamefont{Xu}}, \bibnamefont{and}
  \bibinfo{author}{\bibfnamefont{C.~M.} \bibnamefont{Marcus}},
  \bibinfo{journal}{Phys. Rev. B} \textbf{\bibinfo{volume}{87}},
  \bibinfo{pages}{241401} (\bibinfo{year}{2013}).

\bibitem[{\citenamefont{Jiang et~al.}(2011)\citenamefont{Jiang, Kitagawa,
  Alicea, Akhmerov, Pekker, Refael, Cirac, Demler, Lukin, and
  Zoller}}]{jiang_majorana_2011}
\bibinfo{author}{\bibfnamefont{L.}~\bibnamefont{Jiang}},
  \bibinfo{author}{\bibfnamefont{T.}~\bibnamefont{Kitagawa}},
  \bibinfo{author}{\bibfnamefont{J.}~\bibnamefont{Alicea}},
  \bibinfo{author}{\bibfnamefont{A.~R.} \bibnamefont{Akhmerov}},
  \bibinfo{author}{\bibfnamefont{D.}~\bibnamefont{Pekker}},
  \bibinfo{author}{\bibfnamefont{G.}~\bibnamefont{Refael}},
  \bibinfo{author}{\bibfnamefont{J.~I.} \bibnamefont{Cirac}},
  \bibinfo{author}{\bibfnamefont{E.}~\bibnamefont{Demler}},
  \bibinfo{author}{\bibfnamefont{M.~D.} \bibnamefont{Lukin}}, \bibnamefont{and}
  \bibinfo{author}{\bibfnamefont{P.}~\bibnamefont{Zoller}},
  \bibinfo{journal}{Phys. Rev. Lett.} \textbf{\bibinfo{volume}{106}},
  \bibinfo{pages}{220402} (\bibinfo{year}{2011}).

\bibitem[{\citenamefont{Liu et~al.}(2012{\natexlab{a}})\citenamefont{Liu, Hao,
  Zhu, and Liu}}]{liu_topological_2012}
\bibinfo{author}{\bibfnamefont{G.}~\bibnamefont{Liu}},
  \bibinfo{author}{\bibfnamefont{N.}~\bibnamefont{Hao}},
  \bibinfo{author}{\bibfnamefont{S.-L.} \bibnamefont{Zhu}}, \bibnamefont{and}
  \bibinfo{author}{\bibfnamefont{W.~M.} \bibnamefont{Liu}},
  \bibinfo{journal}{Phys. Rev. A} \textbf{\bibinfo{volume}{86}},
  \bibinfo{pages}{013639} (\bibinfo{year}{2012}{\natexlab{a}}).

\bibitem[{\citenamefont{Liu et~al.}(2013)\citenamefont{Liu, Levchenko, and
  Baranger}}]{liu_floquet_2013}
\bibinfo{author}{\bibfnamefont{D.~E.} \bibnamefont{Liu}},
  \bibinfo{author}{\bibfnamefont{A.}~\bibnamefont{Levchenko}},
  \bibnamefont{and} \bibinfo{author}{\bibfnamefont{H.~U.}
  \bibnamefont{Baranger}}, \bibinfo{journal}{Phys. Rev. Lett.}
  \textbf{\bibinfo{volume}{111}}, \bibinfo{pages}{047002}
  (\bibinfo{year}{2013}).

\bibitem[{\citenamefont{Kundu and Seradjeh}(2013)}]{kundu_transport_2013}
\bibinfo{author}{\bibfnamefont{A.}~\bibnamefont{Kundu}} \bibnamefont{and}
  \bibinfo{author}{\bibfnamefont{B.}~\bibnamefont{Seradjeh}},
  \bibinfo{journal}{Phys. Rev. Lett.} \textbf{\bibinfo{volume}{111}},
  \bibinfo{pages}{136402} (\bibinfo{year}{2013}).

\bibitem[{\citenamefont{Pikulin et~al.}(2012)\citenamefont{Pikulin, Dahlhaus,
  Wimmer, Schomerus, and Beenakker}}]{pikulin_zero-voltage_2012}
\bibinfo{author}{\bibfnamefont{D.~I.} \bibnamefont{Pikulin}},
  \bibinfo{author}{\bibfnamefont{J.~P.} \bibnamefont{Dahlhaus}},
  \bibinfo{author}{\bibfnamefont{M.}~\bibnamefont{Wimmer}},
  \bibinfo{author}{\bibfnamefont{H.}~\bibnamefont{Schomerus}},
  \bibnamefont{and} \bibinfo{author}{\bibfnamefont{C.~W.~J.}
  \bibnamefont{Beenakker}}, \bibinfo{journal}{New J. Phys.}
  \textbf{\bibinfo{volume}{14}}, \bibinfo{pages}{125011}
  (\bibinfo{year}{2012}), .

\bibitem[{\citenamefont{Lee et~al.}(2012)\citenamefont{Lee, Jiang, Aguado,
  Katsaros, Lieber, and De~Franceschi}}]{lee_zero-bias_2012}
\bibinfo{author}{\bibfnamefont{E.~J.~H.} \bibnamefont{Lee}},
  \bibinfo{author}{\bibfnamefont{X.}~\bibnamefont{Jiang}},
  \bibinfo{author}{\bibfnamefont{R.}~\bibnamefont{Aguado}},
  \bibinfo{author}{\bibfnamefont{G.}~\bibnamefont{Katsaros}},
  \bibinfo{author}{\bibfnamefont{C.~M.} \bibnamefont{Lieber}},
  \bibnamefont{and}
  \bibinfo{author}{\bibfnamefont{S.}~\bibnamefont{De~Franceschi}},
  \bibinfo{journal}{Phys. Rev. Lett.} \textbf{\bibinfo{volume}{109}},
  \bibinfo{pages}{186802} (\bibinfo{year}{2012}).

\bibitem[{\citenamefont{Chang et~al.}(2013)\citenamefont{Chang, Manucharyan,
  Jespersen, Nyg\aa{}rd, and Marcus}}]{chang_tunneling_2013}
\bibinfo{author}{\bibfnamefont{W.}~\bibnamefont{Chang}},
  \bibinfo{author}{\bibfnamefont{V.~E.} \bibnamefont{Manucharyan}},
  \bibinfo{author}{\bibfnamefont{T.~S.} \bibnamefont{Jespersen}},
  \bibinfo{author}{\bibfnamefont{J.}~\bibnamefont{Nyg\aa{}rd}},
  \bibnamefont{and} \bibinfo{author}{\bibfnamefont{C.~M.}
  \bibnamefont{Marcus}}, \bibinfo{journal}{Phys. Rev. Lett.}
  \textbf{\bibinfo{volume}{110}}, \bibinfo{pages}{217005}
  (\bibinfo{year}{2013}).

\bibitem[{\citenamefont{Cheng et~al.}(2014)\citenamefont{Cheng, Becker, Bauer,
  and Lutchyn}}]{cheng_interplay_2014}
\bibinfo{author}{\bibfnamefont{M.}~\bibnamefont{Cheng}},
  \bibinfo{author}{\bibfnamefont{M.}~\bibnamefont{Becker}},
  \bibinfo{author}{\bibfnamefont{B.}~\bibnamefont{Bauer}}, \bibnamefont{and}
  \bibinfo{author}{\bibfnamefont{R.~M.} \bibnamefont{Lutchyn}},
  \bibinfo{journal}{Phys. Rev. X} \textbf{\bibinfo{volume}{4}},
  \bibinfo{pages}{031051} (\bibinfo{year}{2014}).

\bibitem[{\citenamefont{\ifmmode~\check{Z}\else \v{Z}\fi{}itko
  et~al.}(2015)\citenamefont{\ifmmode~\check{Z}\else \v{Z}\fi{}itko, Lim,
  L\'opez, and Aguado}}]{zitko_shiba_2015}
\bibinfo{author}{\bibfnamefont{R.}~\bibnamefont{\ifmmode~\check{Z}\else
  \v{Z}\fi{}itko}}, \bibinfo{author}{\bibfnamefont{J.~S.} \bibnamefont{Lim}},
  \bibinfo{author}{\bibfnamefont{R.}~\bibnamefont{L\'opez}}, \bibnamefont{and}
  \bibinfo{author}{\bibfnamefont{R.}~\bibnamefont{Aguado}},
  \bibinfo{journal}{Phys. Rev. B} \textbf{\bibinfo{volume}{91}},
  \bibinfo{pages}{045441} (\bibinfo{year}{2015}).

\bibitem[{\citenamefont{Liu et~al.}(2012{\natexlab{b}})\citenamefont{Liu,
  Potter, Law, and Lee}}]{liu_zero-bias_2012}
\bibinfo{author}{\bibfnamefont{J.}~\bibnamefont{Liu}},
  \bibinfo{author}{\bibfnamefont{A.~C.} \bibnamefont{Potter}},
  \bibinfo{author}{\bibfnamefont{K.~T.} \bibnamefont{Law}}, \bibnamefont{and}
  \bibinfo{author}{\bibfnamefont{P.~A.} \bibnamefont{Lee}},
  \bibinfo{journal}{Phys. Rev. Lett.} \textbf{\bibinfo{volume}{109}},
  \bibinfo{pages}{267002} (\bibinfo{year}{2012}{\natexlab{b}}).

\bibitem[{\citenamefont{Lutchyn et~al.}(2011)\citenamefont{Lutchyn, Stanescu,
  and Das~Sarma}}]{lutchyn_search_2011}
\bibinfo{author}{\bibfnamefont{R.~M.} \bibnamefont{Lutchyn}},
  \bibinfo{author}{\bibfnamefont{T.~D.} \bibnamefont{Stanescu}},
  \bibnamefont{and}
  \bibinfo{author}{\bibfnamefont{S.}~\bibnamefont{Das~Sarma}},
  \bibinfo{journal}{Phys. Rev. Lett.} \textbf{\bibinfo{volume}{106}},
  \bibinfo{pages}{127001} (\bibinfo{year}{2011}).

\bibitem[{\citenamefont{Kells et~al.}(2012{\natexlab{a}})\citenamefont{Kells,
  Meidan, and Brouwer}}]{kells_low-energy_2012}
\bibinfo{author}{\bibfnamefont{G.}~\bibnamefont{Kells}},
  \bibinfo{author}{\bibfnamefont{D.}~\bibnamefont{Meidan}}, \bibnamefont{and}
  \bibinfo{author}{\bibfnamefont{P.~W.} \bibnamefont{Brouwer}},
  \bibinfo{journal}{Phys. Rev. B} \textbf{\bibinfo{volume}{85}},
  \bibinfo{pages}{060507} (\bibinfo{year}{2012}{\natexlab{a}}).

\bibitem[{\citenamefont{Lee et~al.}(2014)\citenamefont{Lee, Jiang, Houzet,
  Aguado, Lieber, and De~Franceschi}}]{lee_spin-resolved_2014}
\bibinfo{author}{\bibfnamefont{E.~J.~H.} \bibnamefont{Lee}},
  \bibinfo{author}{\bibfnamefont{X.}~\bibnamefont{Jiang}},
  \bibinfo{author}{\bibfnamefont{M.}~\bibnamefont{Houzet}},
  \bibinfo{author}{\bibfnamefont{R.}~\bibnamefont{Aguado}},
  \bibinfo{author}{\bibfnamefont{C.~M.} \bibnamefont{Lieber}},
  \bibnamefont{and}
  \bibinfo{author}{\bibfnamefont{S.}~\bibnamefont{De~Franceschi}},
  \bibinfo{journal}{Nat Nano} \textbf{\bibinfo{volume}{9}}, \bibinfo{pages}{79}
  (\bibinfo{year}{2014}), .

\bibitem[{\citenamefont{Rieder et~al.}(2012)\citenamefont{Rieder, Kells,
  Duckheim, Meidan, and Brouwer}}]{rieder_endstates_2012}
\bibinfo{author}{\bibfnamefont{M.-T.} \bibnamefont{Rieder}},
  \bibinfo{author}{\bibfnamefont{G.}~\bibnamefont{Kells}},
  \bibinfo{author}{\bibfnamefont{M.}~\bibnamefont{Duckheim}},
  \bibinfo{author}{\bibfnamefont{D.}~\bibnamefont{Meidan}}, \bibnamefont{and}
  \bibinfo{author}{\bibfnamefont{P.~W.} \bibnamefont{Brouwer}},
  \bibinfo{journal}{Phys. Rev. B} \textbf{\bibinfo{volume}{86}},
  \bibinfo{pages}{125423} (\bibinfo{year}{2012}).

\bibitem[{\citenamefont{Stanescu and Tewari}(2014)}]{stanescu_nonlocality_2014}
\bibinfo{author}{\bibfnamefont{T.~D.} \bibnamefont{Stanescu}} \bibnamefont{and}
  \bibinfo{author}{\bibfnamefont{S.}~\bibnamefont{Tewari}},
  \bibinfo{journal}{Phys. Rev. B} \textbf{\bibinfo{volume}{89}},
  \bibinfo{pages}{220507} (\bibinfo{year}{2014}).

\bibitem[{\citenamefont{Kells et~al.}(2012{\natexlab{b}})\citenamefont{Kells,
  Meidan, and Brouwer}}]{kells_near-zero-energy_2012}
\bibinfo{author}{\bibfnamefont{G.}~\bibnamefont{Kells}},
  \bibinfo{author}{\bibfnamefont{D.}~\bibnamefont{Meidan}}, \bibnamefont{and}
  \bibinfo{author}{\bibfnamefont{P.~W.} \bibnamefont{Brouwer}},
  \bibinfo{journal}{Phys. Rev. B} \textbf{\bibinfo{volume}{86}},
  \bibinfo{pages}{100503} (\bibinfo{year}{2012}{\natexlab{b}}).

\bibitem[{\citenamefont{Roy et~al.}(2013)\citenamefont{Roy, Bondyopadhaya, and
  Tewari}}]{roy_topologically_2013}
\bibinfo{author}{\bibfnamefont{D.}~\bibnamefont{Roy}},
  \bibinfo{author}{\bibfnamefont{N.}~\bibnamefont{Bondyopadhaya}},
  \bibnamefont{and} \bibinfo{author}{\bibfnamefont{S.}~\bibnamefont{Tewari}},
  \bibinfo{journal}{Phys. Rev. B} \textbf{\bibinfo{volume}{88}},
  \bibinfo{pages}{020502} (\bibinfo{year}{2013}).

\bibitem[{\citenamefont{Beenakker}(2014)}]{beenakker_random-matrix_2014}
\bibinfo{author}{\bibfnamefont{C.~W.~J.} \bibnamefont{Beenakker}},
  \bibinfo{journal}{arXiv:1407.2131 [cond-mat]}  (\bibinfo{year}{2014}),
  \bibinfo{note}{arXiv: 1407.2131}.

\bibitem[{\citenamefont{Law et~al.}(2009)\citenamefont{Law, Lee, and
  Ng}}]{law_majorana_2009}
\bibinfo{author}{\bibfnamefont{K.~T.} \bibnamefont{Law}},
  \bibinfo{author}{\bibfnamefont{P.~A.} \bibnamefont{Lee}}, \bibnamefont{and}
  \bibinfo{author}{\bibfnamefont{T.~K.} \bibnamefont{Ng}},
  \bibinfo{journal}{Phys. Rev. Lett.} \textbf{\bibinfo{volume}{103}},
  \bibinfo{pages}{237001} (\bibinfo{year}{2009}).

\bibitem[{\citenamefont{Gaury et~al.}(2014)\citenamefont{Gaury, Weston, Santin,
  Houzet, Groth, and Waintal}}]{gaury_numerical_2014}
\bibinfo{author}{\bibfnamefont{B.}~\bibnamefont{Gaury}},
  \bibinfo{author}{\bibfnamefont{J.}~\bibnamefont{Weston}},
  \bibinfo{author}{\bibfnamefont{M.}~\bibnamefont{Santin}},
  \bibinfo{author}{\bibfnamefont{M.}~\bibnamefont{Houzet}},
  \bibinfo{author}{\bibfnamefont{C.}~\bibnamefont{Groth}}, \bibnamefont{and}
  \bibinfo{author}{\bibfnamefont{X.}~\bibnamefont{Waintal}},
  \bibinfo{journal}{Physics Reports} \textbf{\bibinfo{volume}{534}},
  \bibinfo{pages}{1} (\bibinfo{year}{2014}), .

\bibitem[{\citenamefont{Gaury and Waintal}(2014)}]{gaury_dynamical_2014}
\bibinfo{author}{\bibfnamefont{B.}~\bibnamefont{Gaury}} \bibnamefont{and}
  \bibinfo{author}{\bibfnamefont{X.}~\bibnamefont{Waintal}},
  \bibinfo{journal}{Nat Commun} \textbf{\bibinfo{volume}{5}}
  (\bibinfo{year}{2014}).

\bibitem[{\citenamefont{Chevallier et~al.}(2013)\citenamefont{Chevallier,
  Simon, and Bena}}]{chevallier_andreev_2013}
\bibinfo{author}{\bibfnamefont{D.}~\bibnamefont{Chevallier}},
  \bibinfo{author}{\bibfnamefont{P.}~\bibnamefont{Simon}}, \bibnamefont{and}
  \bibinfo{author}{\bibfnamefont{C.}~\bibnamefont{Bena}},
  \bibinfo{journal}{Phys. Rev. B} \textbf{\bibinfo{volume}{88}},
  \bibinfo{pages}{165401} (\bibinfo{year}{2013}).

\bibitem[{\citenamefont{Groth et~al.}(2014)\citenamefont{Groth, Wimmer,
  Akhmerov, and Waintal}}]{groth_kwant:_2014}
\bibinfo{author}{\bibfnamefont{C.~W.} \bibnamefont{Groth}},
  \bibinfo{author}{\bibfnamefont{M.}~\bibnamefont{Wimmer}},
  \bibinfo{author}{\bibfnamefont{A.~R.} \bibnamefont{Akhmerov}},
  \bibnamefont{and} \bibinfo{author}{\bibfnamefont{X.}~\bibnamefont{Waintal}},
  \bibinfo{journal}{New J. Phys.} \textbf{\bibinfo{volume}{16}},
  \bibinfo{pages}{063065} (\bibinfo{year}{2014}), 
  .

\end{thebibliography}

\end{document}